 \documentclass[twocolumn,trackchanges]{aastex631}
\shorttitle{AASTeX v6.3.1 Sample article}
\shortauthors{Banik \& Ghosh}
\graphicspath{{./}{figures/}}
\begin{document}
%\title{Interpreting observed ultrahigh energy gamma-rays and neutrino emission from the Cygnus Cocoon}
%\title{Investigating the Cygnus Cocoon as a galactic cosmic ray generator using ultrahigh energy gamma-rays and neutrinos}
%\title{Investigating Cygnus Cocoon as a galactic cosmic ray generator using correlated ultrahigh-energy gamma-ray and neutrino observations }
%\title{Investigating Cygnus Cocoon as galactic cosmic ray generator using observed ultrahigh-energy gamma-rays and neutrinos}
%\title{Probing Cygnus Cocoon as galactic cosmic ray generator using correlated ultrahigh-energy gamma-ray and neutrino observations}
\title{Probing the origin of cosmic rays in Cygnus Cocoon using ultrahigh-energy gamma-ray and neutrino observations}
\author{Prabir Banik \thanks{Email address: pbanik74@yahoo.com}}
\affiliation{Department of Physics, $\&$ Center for Astroparticle Physics $\&$ Space Science, \\
Bose Institute, EN-80, Sector-5, Bidhan Nagar,  Kolkata-700091, India}

\author{Sanjay K. Ghosh \thanks{Email address: sanjay@jcbose.ac.in}}
\affiliation{Department of Physics, $\&$ Center for Astroparticle Physics $\&$ Space Science, \\
Bose Institute, EN-80, Sector-5, Bidhan Nagar,  Kolkata-700091, India}

%\collaboration{6}{(AAS Journals Data Editors)}

%\author{Butler Burton}
%\affiliation{Leiden University}
%\affiliation{AAS Journals Associate Editor-in-Chief}

%\altaffiliation{AASTeX v6+ programmer}

%\affiliation{AAS Director of Publishing}
%\affiliation{American Astronomical Society \\
%1667 K Street NW, Suite 800 \\

%\author{Magaret Donnelly}
%\affiliation{IOP Publishing, Washington, DC 20005}



\begin{abstract}
Recent ultrahigh energy gamma-ray observations by the HAWC up to 100 TeV and LHAASO observatories up to 1.4 PeV energies from the direction of Fermi-LAT 4FGL source 4FGL J2028.6+4110e (Cygnus Cocoon), are indicative of a hadronic origin over a leptonic process for their creation. The IceCube Neutrino Observatory has reported IceCube-201120A, a neutrino event coming from the same direction, suggesting that the Cygnus Cocoon may correspond to one of the most plausible sources of high-energy cosmic rays. The apparent relationship of the neutrino event with the observed ultra high energy gamma-rays from Cygnus Cocoon is investigated in this work to study if it can be explained consistently in hadronic interactions of accelerated cosmic rays with ambient matter. Our findings reveal that leptonic mechanisms, together with pure hadronic mechanisms, make a considerable contribution to the understanding of the total electromagnetic spectrum as well as the observed neutrino event. The estimate of expected muon neutrino events from the Cygnus cocoon agrees with the one muon neutrino event detected so far in IceCube multi-year observations. Thus, our results are indicative of the potential of the Cygnus Cocoon to be a galactic cosmic ray source capable of accelerating at least up to PeV energies.

\end{abstract}

\keywords{Cosmic rays --- neutrinos --- gamma-rays}

\section{Introduction} \label{sec:intro}
The Milky Way is known to accelerate cosmic rays with energies up to a few PeV (PeVatrons), however the origin of Galactic cosmic rays has yet to be proven \citep{Hillas83,Berezinskii90}. Supernova remnants (SNRs) are commonly regarded as the most likely origins of Galactic cosmic rays since they are powerful and abundant enough to sustain the intensity of observed cosmic rays \citep{Baade34,Blasi13}. However, normal conditions make it difficult for SNR to accelerate particles to PeV energies \citep{Bell13,Aharonian13}. Moreover, there is just no observational evidence to support SNRs as sources of hadrons with energy more than a few tens of TeV \citep{Helder12,Aharonian19}. Supernova explosions tend to cluster in space (within a few parsecs) and time because large OB stars (the progenitors of core-collapse supernovae) are formed in clusters and live short lives (within a few $10^{5}$ yr) \citep{Higdon05}. As a result, Galactic cosmic rays with energy up to PeV are anticipated to be accelerated by overlapping shocks from SNRs and massive stellar winds (referred to as superbubbles \citep{Tenorio88}) generated around OB associations \citep{Bykov92,Parizot04}. The energy spectra and radial distribution of the calculated cosmic ray flux give evidence for particles accelerated to near-PeV energies in large star clusters \citep{Aharonian19}.

The accelerated cosmic rays in the superbubbles may interact with ambient matter and radiation within the source, producing ultrahigh energy gamma-rays and neutrinos with energies up to PeV. Observations of neutrinos produced in association with ultrahigh energy gamma-rays would unambiguously identify superbubbles as a Galactic cosmic ray PeVatrons. The IceCube collaboration recently announced a candidate track-like neutrino event with an estimated energy of 154 TeV through the standard BRONZE alert procedure \citep{Blaufuss19} on November 20, 2020 \citep{IceCube20,Dzhappuev21}. The neutrino event is likely linked to an extended gamma-ray source Cygnus Cocoon which is a superbubble surrounding a region of OB2 massive star formation \citep{IceCube20}. The Carpet$-$2 experiment reported a $3.1\sigma$ (post-trial) excess of atmospheric air showers from the same direction, consistent with a few months flare in photons over 300 TeV, in temporal correlation with the neutrino event \citep{Dzhappuev21}. Implication of this observation has been discussed in detail in results section. This is the first evidence for a neutrino event being correlated with a Galactic object, despite being determined with considerable uncertainty \citep{Dzhappuev21}.

The Fermi Gamma-Ray Space Telescope's Large Region Telescope (LAT) first identified Cygnus Cocoon emitting hard, multi-GeV gamma-rays in the nearby star-forming area known as Cygnus X \citep{Ackermann11}. The ARGO experiment first identified it at TeV energies \citep{Bartoli14}. The High Altitude Water Cherenkov (HAWC) observatory has reported observations of $1-100$ TeV gamma-rays originating from the Cygnus Cocoon, which may be represented by a power-law below $10$ TeV and exhibits spectrum softening around $10$ TeV \citep{Abeysekara20,Abeysekara21}. The LHAASO collaboration recently reported the discovery of ultrahigh energy photons with energies up to 1.4 PeV from this location, indicating that the spectrum can extend up to $\sim$ 1 PeV \citep{Cao21,Li21}. \cite{Amenomori21} revealed that gamma-ray sources in the Cygnus area contribute significantly to the Galactic-plane diffuse gamma radiation above 400 TeV. These findings imply that the Cygnus Cocoon may be acting as galactic hadronic PeVatron, and could provide significant evidence for understanding the knee of the observed cosmic ray energy spectrum.

%This observation gives support to the apparent link between the neutrino event and the Cygnus Cocoon. 
%The Carpet-2 experiment detected an abundance of gamma-ray events from the same direction, which coincided with the IceCube neutrino signal from the same direction \cite{Dzhappuev21}. The surplus has a statistical significance of $3.1\sigma$ post-trial, immediately supports genuineness of the observed coincidence of the neutrino    \cite{Dzhappuev21}. 

% theoritical models
The star-forming regions such as Cygnus constellation have been proposed as potential sites for cosmic-ray acceleration, as well as gamma-ray and neutrino production \citep{Yang19}. It has been suggested that the high-energy neutrino flux from Cygnus Cocoon will be close to the IceCube sensitivity \citep{Yoast17}. The detected gamma-ray flux might be both leptonic and hadronic in nature. 
%Normally, distinguishing between the leptonic and hadronic origins of observed gamma-ray radiation is quite difficult. 
The HAWC collaboration reported that the observed 1-100 TeV gamma-rays from Cygnus Cocoon were unlikely to be explained by a single electron population emitting gamma-rays from GeV to the highest energy via inverse-Compton emission without its synchrotron radiation exceeding the flux limits set by radio and X-ray studies \citep{Abeysekara21}. Therefore, a significant contribution in the gamma-ray spectrum above a few TeV likely to have hadronic origin for their production. However, the appearance of a cut-off or a break in the measured gamma-ray spectrum at a few TeV is thought to be due to either cosmic ray leakage from the Cocoon or a cut-off in the cosmic ray spectrum injected from the source \citep{Abeysekara21}. Again, the most recent LHASSO observations of ultrahigh-energy gamma-rays up to 1.4 PeV from the Cygnus Cocoon significantly disfavored a leptonic origin for their formation and strongly imply acceleration of cosmic rays at energies greater than PeV \citep{Cao21}. Therefore, the apparent association of the observed neutrino event IceCube-201120A with the measured ultra high energy gamma-rays up to highest energies demands new explanation for their productions. 

Under such circumstances, we would like to examine the apparent link of the neutrino event IceCube-201120A with the observed ultra high energy gamma-rays from Cygnus Cocoon to study whether it can be explained consistently in the framework of hadronic interactions of accelerated cosmic rays with ambient matter. We will also investigate the possibility of a leptonic origin contribution to the total gamma-ray spectra, in addition to a hadronic origin, taking into account IceCube's non-detection of multiple neutrino events. We would also like to inspect the maximum energy that a cosmic ray particle can achieve in the Cygnus Cocoon.

% structure of the paper
The following is the article's structure:
The next section describes the method for evaluating gamma-ray and neutrino fluxes from Cygnus Cocoon.
Section 3 shows numerical estimates of the fluxes of multi-wavelength electromagnetic (EM) spectral energy distribution (SED) and high-energy neutrinos produced by the Cygnus Cocoon. The results are also discussed in the same section. Finally, we will conclude in the section 4.

\section{Methodology}
It can be generally assumed that the electrons and protons are co-accelerated via the diffusive shock acceleration mechanism in the interacting winds created by the collective activity of massive stars in the Cygnus Cocoon. We can assume that a fraction of the stellar wind energy, $\eta_e$, may be used to accelerate electrons and a fraction, $\eta_p$, can be used to accelerate hadrons \citep{Bednarek07}.

In this case, we consider a broken power-law energy distribution of shock-accelerated electrons with spectral indices $\alpha_1$ and $\alpha_2$ before and after the spectral break at Lorentz factor $\gamma_b$, as illustrated below \citep{Katarzynski01}

\begin{eqnarray}
N_e(\gamma_e) = K_e \gamma_e^{-\alpha_1} \hspace{1.5cm} \mbox{if}\hspace{0.6cm} \gamma_{e,min} \le \gamma_e \le \gamma_b \nonumber \\
         = K_e \gamma_b^{\alpha_2-\alpha_1} \gamma_e^{-\alpha_2} \hspace{0.42cm} \mbox{if}\hspace{0.48cm} \gamma_b <\gamma_e \le \gamma_{e,max}\;
\label{Eq:1}
\end{eqnarray}
where $K_e$ represents the normalization constant which is related with the available stellar wind power ($L_w$) as given bellow
\begin{eqnarray}
\eta_e \frac{L_{w}}{4 \pi r^2 v_w} = m_e c^2 \displaystyle\int_{\gamma_{e,min}}^{\gamma_{e,max}}\gamma_{e} N_e(\gamma_e) d\gamma_{e}.
\label{Lcr}
\end{eqnarray}
where $\gamma_e = E_e/m_e c^2$ represents the Lorentz factor of electrons of energy $E_e$, $r$ is the radius of the Cygnus Cocoon, and $v_w$ is a typical stellar wind velocity. For a radiative cooling break in a uniform magnetic field $B$, the electron distribution breaks in its index by one power (i.e, $\Delta \alpha = \alpha_2-\alpha_1 \approx 1$) above the spectral break Lorentz factor $\gamma_b'$ \citep{Longair94}. 
%For cooling break, the electron break Lorentz factor $\gamma_b$ can be estimated by equalizing the radiative cooling timescale $\tau_{cool} = \frac{3m c}{4(u_B+u_{ph})\sigma_T \gamma_e}$ with the diffusion timescale $t_{diff} = \frac{r^2}{D_{diff}}$, where $\sigma_T$ is the Thomson scattering cross section, $u_{B} = \frac{B^2}{8\pi}$ represents the magnetic field energy density corresponds to magnetic field $B$, $u_{ph}$ represents the soft photon energy density including synchrotron, star light and dust emission within the Cocoon, and $D_{diff} = D_0 (\frac{E_e}{1\,GeV})^{0.33}$ \citep{Abeysekara21} denotes the diffusion coefficient of accelerated electrons. The maximum energy of electrons can be computed by equating synchrotron cooling time with acceleration timescale of electrons and can be written as $E_{e,max} = 58 (\frac{\xi_{-5}}{B_{-5}})^{1/2}$ TeV where $\xi_{e} = 10^{-5} \xi_{-5}$ represents the so-called acceleration coefficient ($\xi_e \le 1$), and $B = 10^{-5} B_{-5}$ is the magnetic field \citep{Bednarek07}. %as given in equation 1 of \cite{Bednarek07}.

The low-energy component of the EM SED extending from radio to x-ray energies generated in the Cygnus Cocoon is represented by synchrotron radiation of primary accelerated electrons, which is estimated here using the methodology given by \cite{Bottcher13}. The inverse Compton (IC) scattering of primary accelerated electrons with target photons contributes significantly to the observed EM spectrum in the MeV to GeV ranges and may be estimated using the formulas presented in \cite{Blumenthal70}. Here, we consider four target radiation fields following \cite{Ackermann11} for gamma-ray generation via IC scattering, including synchrotron photons, strong stellar light fields around Cyg OB2 and NGC 6910 (a star cluster in the neighborhood of OB2), and a more diffuse dust radiation field over the whole Cocoon. The Bremsstrahlung scattering of primary accelerated electrons with ambient matter of density $n_H$ is found to explain the observed gamma-ray spectrum in GeV to few TeV energy band and may be estimated by following \cite{Blumenthal70}. 

The cosmic ray (protons) production spectrum is also expected to follow a power law \citep{Malkov01}  
\begin{eqnarray}
\frac{dn_p}{dE_p} = K_p E_p^{-\alpha_p}
\label{cr_prod}
\end{eqnarray}
where $E_p$ denotes the energy of the cosmic ray proton, $\alpha_p$ is the spectral index, and the proportionality constant is $K_p$, which may be derived by using the fraction of stellar wind energy carried by cosmic ray protons as follows
\begin{eqnarray}
\eta_p \frac{L_{w}}{4 \pi r^2 v_w} =  \displaystyle\int_{E_{p,min}}^{E_{p,max}}E_p \frac{dn_p}{dE_p} dE_p
\label{Lcr}
\end{eqnarray}
where $E_{p,min}$ and $E_{p,max}$ denotes the minimum and maximum energies of accelerated cosmic ray protons, respectively. 
%The maximum energy ($E_{p,max}$) may be restricted when the acceleration timescale $t_{acc} = \frac{E_p}{\xi_p e B c}$ matches the lifetime ($t_{age} \sim 2$ Myrs) of the Cocoon, where $\xi_p$ ($\leq 1$) is the acceleration coefficient. 

The interaction of shock-accelerated cosmic ray protons with ambient matter (protons) of density $n_H$ can explain the high energy gamma-ray emission from the Cygnus Cocoon. Such hadronic ($pp$) interaction produces neutral and charged pions, which decay to create high-energy gamma-rays and neutrinos, respectively. To estimate the high energy gamma rays and neutrino emissivities ($Q_{\gamma}^{pp}$ \& $Q_{\nu}^{pp}$ respectively) produced in hadronic interactions ($pp$) in the Cygnus Cocoon, we follow Refs.~\citep{Kelner06,Banik17,Banik19}. The associated differential flux of high-energy gamma-rays and muon neutrinos reaching Earth from the Cygnus Cocoon may be represented as
%To estimate the differential flux of high energy gamma rays ($\frac{d\Phi_{\gamma}}{dE_{\gamma}}$) reaching the Earth from the Cygnus Cocoon, we follow Refs.~\citep{Kelner06,Banik17}. The associated flux of high-energy muon neutrinos reaching Earth from the Cygnus Cocoon may be represented as 
\begin{eqnarray}
%\small
\frac{d\Phi_{\gamma/\nu_{\mu}}}{dE_{\gamma/\nu}} = \zeta \frac{V}{4\pi d^{2}} Q_{\gamma/\nu}^{pp}(E_{\gamma/\nu})
\end{eqnarray}
where $\zeta$ is a constant equal to $1$ for gamma rays and equal to $1/3$ for muon neutrinos due to neutrino oscillation, $V =\frac{4}{3}\pi r^3$ represents the volume of the emission region, and $d$ is the distance between the Cygnus Cocoon and the Earth. The number of expected muon neutrino events in the IceCube detector in time $t$ may be computed using the following formula
\begin{eqnarray}
N_{\nu_{\mu}} = t \int_{E_{\nu,min}}^{E_{\nu,max}} A_{eff}(E_{\nu}). \frac{d\Phi_{\nu_{\mu}}}{dE_{\nu}} dE_{\nu}
\label{event}
\end{eqnarray}
where $A_{eff}$ is the effective area of the IceCube detector (IC86) at the declination of the source \citep{Icecube21}. We can choose $E_{\nu,min} \approx 30$ TeV which is in good agreement with the effective energy threshold of the IceCube detector for astrophysical neutrinos \citep{Taboada16}.

\section{Results and Discussion}
The Cygnus Cocoon is situated at RA $307.17$ deg and Dec $41.17$ deg (J2000) \citep{IceCube20}. It has an angular size of around $2.1^{\circ}$, corresponding to a radius of $r = 55$ pc at a distance of $d = 1.4$ kpc from Earth \citep{Abeysekara21}. The Cygnus Cocoon comprises of two star clusters, Cyg OB2 and NGC 6910, with total wind power estimates of ($2-3$) $\times 10^{38}$ erg/s and ($1-1.5$) $\times 10^{36}$ erg/s, respectively \citep{Ackermann11}. Here, we consider a typical stellar wind with velocity $v_w = 10^3$ km/s \citep{Ackermann11} and wind power $L_w = 3\times 10^{38}$ erg/s to estimate stellar wind energy density. 

The IceCube neutrino observatory has reported a neutrino event IceCube-201120A, which is likely to be associated to Cygnus Cocoon. The Carpet$-$2 experiment observed an excess of gamma-ray events consistent with a few months flare in photons above 300 TeV from the direction of the Cygnus region, in temporal and spatial coincidence with the IceCube neutrino alert \citep{Dzhappuev21}. As the gamma-ray emission from the cocoon (the radius $\sim 55$ pc) cannot vary on a time scale of months, this neutrino event is more likely to be due to any compact source within the cocoon rather than the diffused emission. Therefore, if this event is truly related with the few months gamma-ray flare, it cannot be used to limit hadronic gamma-ray emission from the cocoon. However, the neutrino event may not be linked to such a flare because no excess GeV-TeV gamma-ray flux was measured from the Cygnus region during the neutrino arrival period by Fermi-Lat and HAWC \citep{Garrappa20,Ayala20}. 
% However, during the neutrino arrival period, Fermi-Lat and HAWC observed no excess GeV-TeV gamma-ray flux from the Cygnus region \citep{Garrappa20,Ayala20}, which does not imply a gamma-ray flare.

Recently, the LHAASO observatory detected gamma-ray flux of $0.54 (0.10)$ CU at 100 TeV (CU, the Crab Nebula flux at 100 TeV; 1 CU $= 6.1 \times 10^{-17}$ photons TeV$^{-1}$  cm$^{-2}$  s$^{-1}$) from the direction of the source LHAASO J2032+4102 (Cygnus Cocoon) when only half of its KM2A detectors were operational \citep{Cao21}. The LHAASO observatory found $45$ on-source events (with 6.7 number of background events) with energies above 100 TeV up to 1.4 PeV during an exposure period of $2648.2$ hr from the direction of Cygnus Cocoon \citep{Cao21}. We estimated the corresponding gamma-ray flux seen by LHAASO from Cygnus Cocoon by assuming a power-law gamma-ray spectrum of $f_{\gamma} = N_0 E^{-\Gamma}$ with a photon index of $\Gamma\approx 2.7$ in the energy range of 100 TeV to 1.4 PeV. %, which is shown as a blue continuous line in Fig.~\ref{Fig:1}. %for the other three sources \cite{Cao21}.
% The comparable gamma-ray flux from Cygnus Cocoon as observed by LHAASO is computed in the energy range of 100 TeV to 1.4 PeV using the relation
The normalization constant $N_0$ of the observed gamma-ray flux from Cygnus Cocoon by LHAASO may be computed as \citep{Aharonian20}
\begin{eqnarray}
S_{\gamma} = \epsilon T_{ex}\int_{100\; TeV}^{1.4 PeV} A_{eff}^\gamma f_{\gamma} dE_{\gamma}
\label{LHAASO}
\end{eqnarray}
where $S_{\gamma}$ represents the number of gamma-ray signal event in LHAASO detector, $A_{eff}^\gamma$ denotes the effective area when only half of its KM2A detectors were operational \citep{Aharonian21}, and $T_{ex}$ represents corresponding exposure time \citep{Cao21} for Cygnus Cocoon. Here, $\epsilon = 0.68$ is the fraction of observed event counts within the angular resolution of the instrument \citep{He19}. The combined gamma-ray spectra from GeV to highest energies ($1.4$ PeV) from Cygnus Cocoon as observed by Fermi-LAT \citep{Ackermann11,Astiasarain21}, ARGO \citep{Bartoli14}, HAWC \citep{Abeysekara21}, and LHAASO \citep{Cao21,Li21} observatories indicate hadronic origin for their generation \citep{Abeysekara21,Cao21}. 

%On the other hand, the observed neutrino event IceCube-201120A is found to be in temporal correlation with the few months of gamma-ray flare over 300 TeV from the direction of Cygnus Cocoon \citep{Dzhappuev21}. However, the gamma-ray emission from the cocoon (the radius $\sim 55$ pc) cannot vary on a time scale of months. This suggests that this neutrino event is rather related to any compact source within the cocoon but not to the diffused emission. Therefore, if this event truly related with the few months gamma-ray flare, it cannot be used to limit hadronic gamma-ray emission from the cocoon. 

The likelihood of a leptonic origin contribution to the overall gamma-ray spectra, in addition to a hadronic origin, is examined below, taking into consideration IceCube's non-detection of multiple neutrino events from Cygnus Cocoon. 

\subsection{Pure hadronic origin}
We have estimated the gamma-ray flux produced in the hadronic interaction ($pp$) of a single population of accelerated cosmic rays (protons) with the ambient protons within the astronomical object Cygnus Cocoon. Because the Cygnus area contains a massive molecular cloud complex with a total mass of $8\times 10^{6} M_{\odot}$ \citep{Ackermann12}, the interstellar gas density should be more than $10$ cm$^{-3}$ \citep{Bartoli14}. To explain the observed EM SED, we choose an ambient matter density of $n_H = 30$ cm$^{-3}$ in the area, as suggested by HI and HII observations \citep{Abeysekara21}. Here, we adopt a magnetic field of $B = 20$ $\mu$G as inferred from pressure balance with the gas throughout the Cygnus Cocoon region \citep{Ackermann11}.

\begin{figure}[h]
  \begin{center}
% \scalebox{2.5}{
  \includegraphics[width = 0.48\textwidth,height = 0.45\textwidth,angle=0]{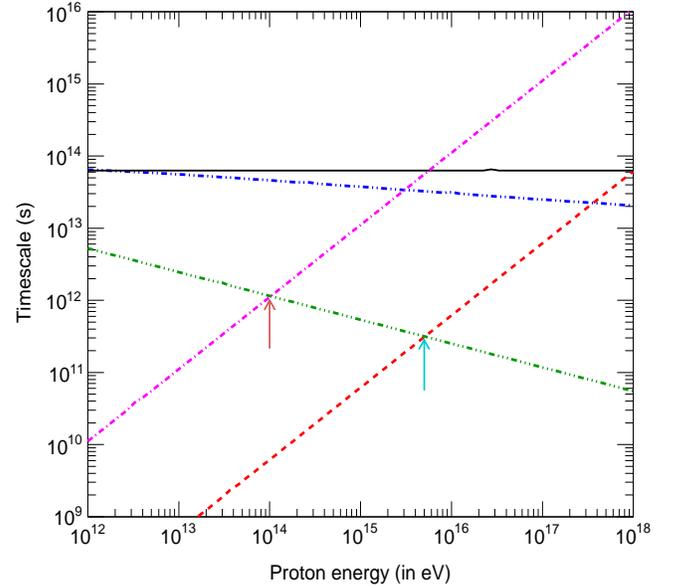}
\end{center}
%  \captionsetup{margin=50pt,font=small,labelfont=bf}
\caption{The estimated relevant timescales for protons. The red dashed and magenta dash-single-dotted lines represent the acceleration timescales with $\xi_p = 9\times 10^{-5}$ and $5\times 10^{-7}$ respectively. The blue dash-double-dotted, and green dash-triple-dotted lines denote the energy loss time-scale for protons in $pp-$interaction and diffusion timescale respectively. The black continuous line indicates the lifetime of the Cocoon. The points, as indicated by the cyan and brown arrows, represent the maximum achievable energy by a cosmic ray proton with $\xi_p = 9\times 10^{-5}$ and $5\times 10^{-7}$ respectively. }
\label{Fig:1}
\end{figure}

We compared the acceleration time-scale of a proton with its energy loss time-scale in $pp-$interaction, diffusion timescale, and the age of the Cocoon ($t_{age} \sim 2$ Myr) to understand the maximum achievable energy by a cosmic ray particle within the Cocoon. The timescale of acceleration of cosmic ray protons can be represented as ($t_{acc} = \frac{E_p}{\xi_p e B c}$), where $\xi_p$ ($\leq 1$) is the proton acceleration coefficient. The energy loss time-scale for protons in $pp-$interaction can be written as $t_{pp} = \frac{1}{k_{pp}\sigma_{pp} n_H c}$, where $k_{pp} = 0.45$ and $\sigma_{pp}$ represent the in-elasticity \citep{Gaisser90} and the interaction cross section \citep{Kelner06}, respectively. The diffusion timescale can be represented by $t_{diff} = \frac{r^2}{D_{diff}}$ \citep{Bednarek07}, where $D_{diff} = D_0 (\frac{E_p}{10\,GeV})^{\delta}$ \citep{Giuliani10,Berezinskii90} denotes the diffusion coefficient of accelerated protons. We may choose $D_{0} = 1.2\times 10^{27}$ cm$^2$/s as the diffusion coefficient at $10$ GeV energy since the dense gaseous medium has a slower diffusion than the galactic medium ($\approx 10^{28}$ cm$^2$/s in our Galactic medium)  \citep{Berezinskii90,Aharonian96}. We have taken into account $\delta = 0.33$, as recently found by the Alpha Magnetic Spectrometer (AMS-02) when measuring the boron to carbon flux ratio in cosmic rays \citep{Aguilar16}. The aforementioned timescales of relativistic protons as functions of proton energy are displayed in Fig.~\ref{Fig:1} (also see \cite{Bednarek07}). By comparing the acceleration time-scale of a proton with its diffusion timescale (e.g. \cite{Bednarek07}), %its energy loss time-scale in $pp-$interaction, 
we found that hadrons can be accelerated up to $5\times 10^{15}$ eV (or $10^{14}$ eV) within the cocoon with an acceleration coefficient of $\xi_p = 9\times 10^{-5}$ (or $5\times 10^{-7}$). 

\begin{figure}[h]
  \begin{center}
% \scalebox{2.5}{
  \includegraphics[width = 0.48\textwidth,height = 0.45\textwidth,angle=0]{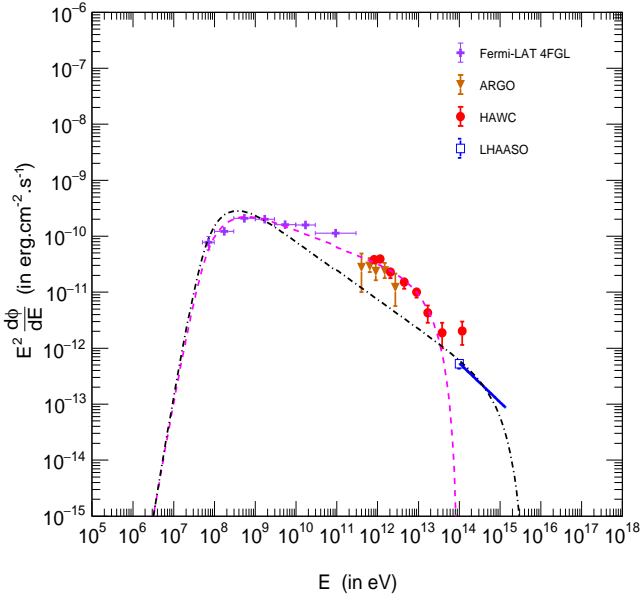}
\end{center}
%  \captionsetup{margin=50pt,font=small,labelfont=bf}
\caption{The estimated differential EM SED reaching at earth from the Cygnus Cocoon in pure hadronic origin scenario. The black dash-dotted and magenta dashed lines represent the EM spectrum produced with ($\eta_{p} = 100\%$, $\alpha_p = -2.6$, and $E_{p,max} = 5\times 10^{15}$ eV) and ($\eta_{p} = 1.6\%$, $\alpha_p = -2.35$, and $E_{p,max} = 10^{14}$ eV), respectively. The blue continuous line represent our estimated gamma-ray flux limit detected by LHAASO.}
\label{Fig:2}
\end{figure}

To match the observed gamma-ray spectrum, we first consider a primary cosmic-ray production spectrum obtained by assuming an acceleration efficiency of $\eta_p = 100\%$ of cosmic ray protons with a spectral index of $\alpha_p = -2.6$, and the maximum energy of $E_{p,max} = 5\times 10^{15}$ eV. According to our findings, an estimated gamma-ray flux based on a single power-law distribution of accelerated cosmic rays cannot properly explain the observed overall gamma-ray spectrum up to energies of 1.4 PeV. This is mostly due to a spectral break in the gamma-ray spectrum around 10 TeV energy, which has been seen in both ARGO and HAWC detector studies. %The associated muon neutrino event in IceCube detector from Cygnus Cocoon is estimated using Eq.~\ref{event} and determined to be $\sim 0.8$ events above $30$ TeV energy in $10$ years. 

When we consider $\eta_{p} = 1.6\%$, i.e. the portion of the stellar wind energy carried by cosmic ray protons with a spectral index of $\alpha_p = -2.35$ and a maximum achievable energy of $E_{p,max} = 10^{14}$ eV, the observed gamma-ray spectrum can be reproduced well from GeV to few tens of TeV energy range. However, it was unable to explain the reported PeV gamma-ray flux by the LHAASO detector. The estimated differential gamma-ray flux reaching the Earth from the Cygnus Cocoon for two stated scenarios are displayed in the Fig.~\ref{Fig:2} along with the observations. %We estimated that $0.04$ muon-neutrino events with energies greater than 30 TeV should be detected in the IceCube detector from Cygnus Cocoon in $10$ years.

In the next section, we have investigated whether viable leptonic processes, in conjunction with pure hadronic mechanisms, can explain the observed spectrum of gamma-rays, as well as the reported neutrino event from Cygnus Cocoon.%, as mentioned below. 

\subsection{Lepto-hadronic origin}
\begin{figure}[h]
  \begin{center}
% \scalebox{2.5}{
  \includegraphics[width = 0.48\textwidth,height = 0.45\textwidth,angle=0]{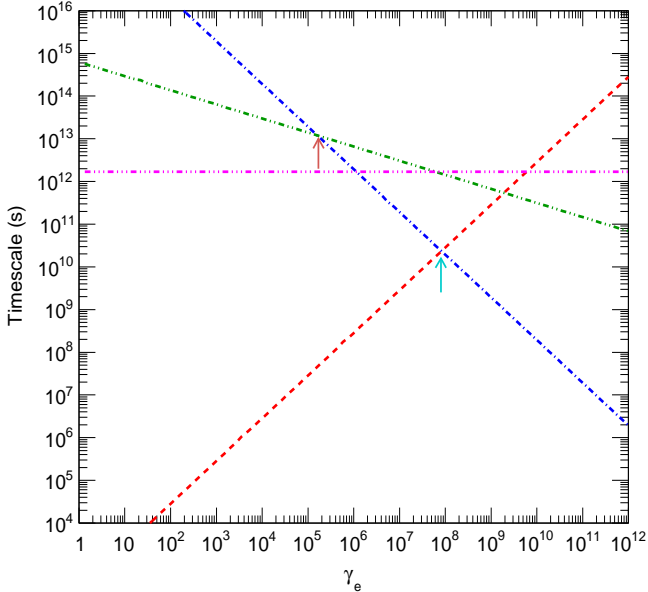}
\end{center}
%  \captionsetup{margin=50pt,font=small,labelfont=bf}
\caption{The estimated relevant timescales for electrons. The red dashed and blue dash-single-dotted lines represent the acceleration timescale of electron with $\xi_e = 10^{-5}$ and synchrotron cooling timescale respectively. The green dash-double-dotted, and magenta dash-triple-dotted lines denote the diffusion, and advection timescales respectively. The points, as indicated by the cyan and brown arrows, represent the maximum achievable Lorentz factor, and spectral break Lorentz factor of a electron, respectively.}
\label{Fig:3}
\end{figure}

The source is thought to accelerate both electrons and protons at the same time. The acceleration timescale of electrons is expressed similarly to that of protons, but with $\xi_e\le 1$ as the acceleration coefficient. To estimate the diffusion timescale, we use $D_{diff} = 1.2\times 10^{27} (\frac{E_e}{10\,GeV})^{0.33}$ cm$^2$/s as the diffusion coefficient for electrons, which is the same as that for protons at 10 GeV energy. The advection timescale can be estimated as $t_{adv} = r/v_w$ \citep{Bednarek07a}. The maximum electron Lorentz factor ($\gamma_{e,max}$) was calculated by matching the acceleration timescale with the synchrotron energy loss timescale ($t_{cool} = \frac{3m c}{4u_B\sigma_T \gamma_e}$) and was found to be $8\times 10^7$ with an acceleration coefficient of $\xi_e = 10^{-5}$. Here, $u_{B} = \frac{B^2}{8\pi}$ represents the magnetic field energy density and $\sigma_T$ is the Thomson scattering cross section. The synchrotron cooling timescale begins to take precedence over the diffusive timescale, i.e. the average time spent by electrons with energy $E_e$ inside the cocoon region, at Lorentz factor $1.7\times 10^5$ as shown in Fig.~\ref{Fig:3}, which can be regarded as the spectral break Lorentz factor $\gamma_b$.  

\begin{figure*}[t]
  \begin{center}
% \scalebox{2.5}{
  \includegraphics[width = 1.0\textwidth,height = 0.45\textwidth,angle=0]{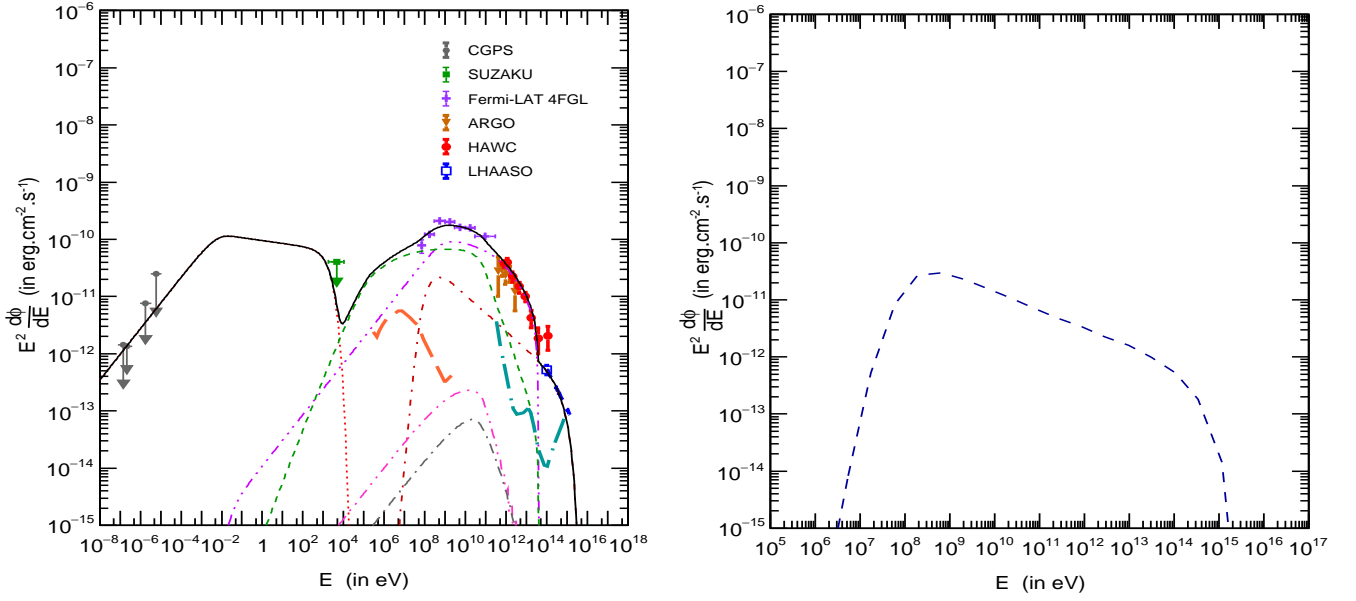}
\end{center}
%  \captionsetup{margin=50pt,font=small,labelfont=bf}

\caption{Left: The estimated differential EM SED reaching earth from the Cygnus Cocoon in the Lepto-hadronic origin scenario. The red dotted and green small-dashed lines represent the EM spectrum produced by synchrotron emission and Bremsstrahlung scattering of primary relativistic electrons in ambient matter, respectively. The EM spectrum produced by IC scattering of relativistic electrons with star light fields around NGC 6910 and Cyg OB2, as well as a dust radiation field, is denoted by the gray long-dash-single-dotted, magenta long-dash-double-dotted, and violet long-dash-triple-dotted lines, respectively. The gamma-ray flux created in pp interactions between relativistic protons and ambient protons is indicated by the brown small-dash-single-dotted line. The black continuous line shows the estimated overall differential multiwavelength EM SED coming from the Cygnus Cocoon. The cyan very-long-dash-single-dotted, and orange very-long-dashed lines indicate the detection sensitivity of the LHAASO, and e-ASTROGAM detectors for one years of observation, respectively. The blue long dashed line represents our estimated gamma-ray flux limit detected by LHAASO. Right: The estimated corresponding all flavor neutrino flux reaching the Earth from the Cygnus Cocoon.}
\label{Fig:4}
\end{figure*}

We consider a broken power-law distribution of accelerated primary electrons with spectral indices of $\alpha_1 = 2.1$ and $\alpha_2 = 3.1$ before and after the spectral break at the Lorentz factor $\gamma_b = 1.7\times 10^5$, derived by assuming a fraction of wind energy $\eta_e = 9\%$ carried by the electrons. %We use $\gamma_{e,max} = 8\times 10^7$ as the maximal electron Lorentz factor, which corresponds to an acceleration coefficient of $\xi_e = 10^{-5}$. 
The synchrotron emission of the accelerated electrons has been computed and compared to multi-wavelength observations of the Cygnus Cocoon from radio to X-ray energies. %The synchrotron emission is estimated using a magnetic field of $B = 20$ $\mu$G inferred from pressure balance with the gas throughout the Cygnus Cocoon region \citep{Ackermann11}. 
The IC scattering of primary relativistic electrons with synchrotron photons co-moving within the source is found to have no substantial contribution to the observed EM spectrum from the source at energies ranging from MeV to TeV. A significant contribution to the observed EM spectrum in the MeV to few tens of TeV energy range is found to be produced by IC scattering of primary relativistic electrons with strong star light fields around Cyg OB2 and NGC 6910 (a star cluster in the neighborhood of OB2), and a more diffuse dust radiation field over the entire Cocoon. Furthermore, we have found that the gamma-ray flux produced by Bremsstrahlung scattering of primary relativistic electrons in ambient matter with a density of $n_H = 30$ cm$^{-3}$ may contribute significantly to the measured gamma-ray energies ranging from MeV to TeV from the source.

The interactions of relativistic primary cosmic rays with ambient protons in the source can contribute significantly to EM SED above 100 TeV energies, as detected by the LHAASO observatory, as well as generate high energy neutrinos. The required fraction of star wind energy carried by the accelerated primary protons is determined to be $\eta_{p} = 8\%$ with the best fitting spectral slope of $\alpha_p = - 2.4$ and the maximum attainable energy $E_{p,max} \simeq 5\times 10^{15}$ eV. The left panel of Fig.~\ref{Fig:4} shows the estimated differential gamma-ray spectra escaping from the Cygnus Cocoon along with the different satellite and ground-based observational data. Using Eq.~\ref{LHAASO} and the model-estimated gamma-ray flux from Cygnus Cocoon, we again computed the predicted gamma-ray signal events in the LHAASO detector from the source and found that they were compatible with the observations. 

The estimated neutrino flux from Cygnus Cocoon reaching at Earth is displayed in the right panel of the Fig.~\ref{Fig:4}. The expected muon neutrino event in the IceCube detector from Cygnus Cocoon is estimated to be roughly $N_{\nu_\mu} = 0.65$ event above 30 TeV energy in $10$ years using Eq.~(\ref{event}). Because a possible contribution to the event rates due to interactions of tau-neutrinos that create muons with a branching ratio of $17.7$\% was not addressed, the estimated number of neutrino events is conservative \citep{Ansoldi18,Banik20}. As a result, the total muon like neutrino events may be computed as $N_{\mu}^{like} = N_{\mu} + 17.7\% \times N_{\mu}$, resulting in a $0.77$ event. Our estimate of expected muon neutrino events is consistent with the only one muon neutrino event reported so far from the cocoon direction in IceCube multi-year observations. The model fitted parameters are displayed in Table~\ref{table1}. 

\begin{table}[t]
  \begin{center}
    \caption{Model fitting parameters for Cygnus Cocoon according to Lepto-hadronic model.}
    \label{table1}
    \begin{tabular}{c|c}
      \toprule
      Parameters &  Values   \\ \hline
       $L_w$ (in erg/s)          &   $3\times 10^{38}$   \\
       $v_w$ (in km/s)          &    $10^3 $   \\
       $B$ (in $\mu G$)          &  $20$  \\
       $\alpha_1$           &   $- 2.1$  \\ 
       $\alpha_2$           &   $- 3.1$  \\
       $\gamma_b$          &  $1.7\times 10^5$ \\
       $\gamma_{e,min}$    &  $1$ \\
       $\gamma_{e,max}$    &  $8\times 10^7$ \\
       $\eta_e$    & $9\%$  \\ 
       $n_H$  (in cm$^{-3}$) &  $30$  \\
       $\alpha_p$           &   $- 2.4$  \\
       $E_{p,min}$ (in eV) &  $10^{9}$  \\ 
       $E_{p,max}$ (in eV) &  $5\times 10^{15}$  \\ 
       $\eta_p$    &  $8\%$  \\  \\ \hline    \hline
    \end{tabular}
  \end{center}
\end{table}

\section{Conclusion}
The observed 1-100 TeV gamma-rays from Cygnus Cocoon is unlikely to be explained by a single electron population emitting gamma-rays from GeV to the highest energy via inverse-Compton, and Bremsstrahlung emission without exceeding the flux limits established by radio and X-ray studies. Our findings show that the combined gamma-ray spectra from GeV to maximum energies ($1.4$ PeV) from Cygnus Cocoon as reported by Fermi-LAT, ARGO, HAWC, and LHAASO observatories cannot be explained by pure hadronic ($pp$) interactions of relativistic cosmic rays with ambient matter. Our results suggest that leptonic processes, in combination with pure hadronic mechanisms, are necessary to consistently represent the complete electromagnetic spectrum. Particularly, the detected gamma-ray flux in sub-PeV energies by LHAASO is found to be best explained by hadronic interaction of cosmic rays, which originated in the Cygnus Cocoon, with ambient matter. The single muon neutrino event detected so far from the cocoon direction in IceCube multi-year data agrees with our estimate of expected muon neutrino events. Thus, the Cygnus Cocoon might be one of the long-suspected galactic PeVatrons, capable of accelerating cosmic rays with energies at-least upto few PeV, providing strong evidence for the origin of knee in the observed cosmic ray energy spectrum. Future gamma ray telescopes with better sensitivity than current generation gamma ray telescopes, such as e-ASTROGAM \citep{Angelis18}, CTA \citep{Ong17} and LHAASO (complete operational mode) \citep{Liu17}, and future neutrino telescope IceCube-Gen2 \citep{Aartsen21}, KM3NeT \citep{Aiello19} with better sensitivities may offer a clearer understanding of the physical origin of gamma rays and neutrino emission.

\begin{acknowledgments}
The authors would like to thank an anonymous reviewer for valuable remarks that helped us improve and correct the manuscript. PB thanks financial support from the SERB (DST), Government of India, under the fellowship reference number PDF/2021/001514.

%We thank all the people that have made this AASTeX what it is today.  This
%includes but not limited to Bob Hanisch, Chris Biemesderfer, Lee Brotzman,
%Pierre Landau, Arthur Ogawa, Maxim Markevitch, Alexey Vikhlinin and Amy
%Hendrickson. Also special thanks to David Hogg and Daniel Foreman-Mackey
%for the new "modern" style design. Considerable help was provided via bug
%reports and hacks from numerous people including Patricio Cubillos, Alex
%Drlica-Wagner, Sean Lake, Michele Bannister, Peter Williams, and Jonathan
%Gagne.
\end{acknowledgments}

\bibliography{AAS38324R2}{}
\bibliographystyle{aasjournal}



\end{document}